\documentclass[prc,twocolumn,,superscriptaddress,showpacs,floatfix]{revtex4}

\usepackage{graphicx}
\usepackage{dcolumn}

\begin{document}               

\def\be{\begin{equation}}
\def\ee{\end{equation}}
\def\ba{\begin{eqnarray}}
\def\ea{\end{eqnarray}}
\def\bas{\begin{eqnarray*}}
\def\eas{\end{eqnarray*}}

%\addtolength{\topmargin}{2cm}

\title{Factorization of shell-model ground-states}

\author{T.~Papenbrock}
\affiliation{Physics Division, 
Oak Ridge National Laboratory, Oak Ridge, TN 37831, USA}
\affiliation{Department of Physics and Astronomy, University of Tennessee,
Knoxville TN 37996-1201, USA}
\author{D.~J.~Dean}
\affiliation{Physics Division, 
Oak Ridge National Laboratory, Oak Ridge, TN 37831, USA}
\date{\today}

\begin{abstract}
We present a new method that accurately approximates the shell-model
ground-state by products of suitable states.  The optimal factors are
determined by a variational principle and result from the solution of
rather low-dimensional eigenvalue problems.  The power of this method
is demonstrated by computations of ground-states and low-lying excitations
in $sd$-shell and $pf$-shell nuclei.
\end{abstract}


\maketitle 

Realistic nuclear structure models are notoriously difficult to solve
due to the complexity of the nucleon-nucleon interaction and the sheer
size of the model spaces. Exact diagonalizations are now possible for
$pf$-shell nuclei \cite{Antoine,Nathan} and for sufficiently light
systems \cite{Navratil00,Navratil02}, and Quantum Monte Carlo calculations
\cite{Pieper01,Pieper02} have solved light nuclei up to mass number $A=10$.
For cases where an exact diagonalization is not feasible, various accurate
approximations are employed. Important examples are stochastic methods
like Shell-Model Monte Carlo \cite{Lang93,Koonin97} and Monte Carlo
Shell-Model \cite{MCSM}. Recently, non-stochastic approximations have
been developed. Examples are the Exponential Convergence Method
\cite{Horoi94,Horoi99,Horoi02,Horoi03} and the application of density matrix
renormalization group to nuclear structure problems
\cite{Duk01,Duk02,Dimitrova02}. These latter methods truncate the
Hilbert space to those states that accurately approximate the
ground-state, and the results of such calculations usually converge
exponentially as one increases the number of retained states. The
correct identification of the most suitable states clearly becomes
crucial for these approximations to be useful. In this work, we propose
a solution to this problem and present a method that obtains the {\it
optimal} states from a variational principle.

\begin{figure}[t]
\vskip 0.3cm
\includegraphics[width=0.45\textwidth]{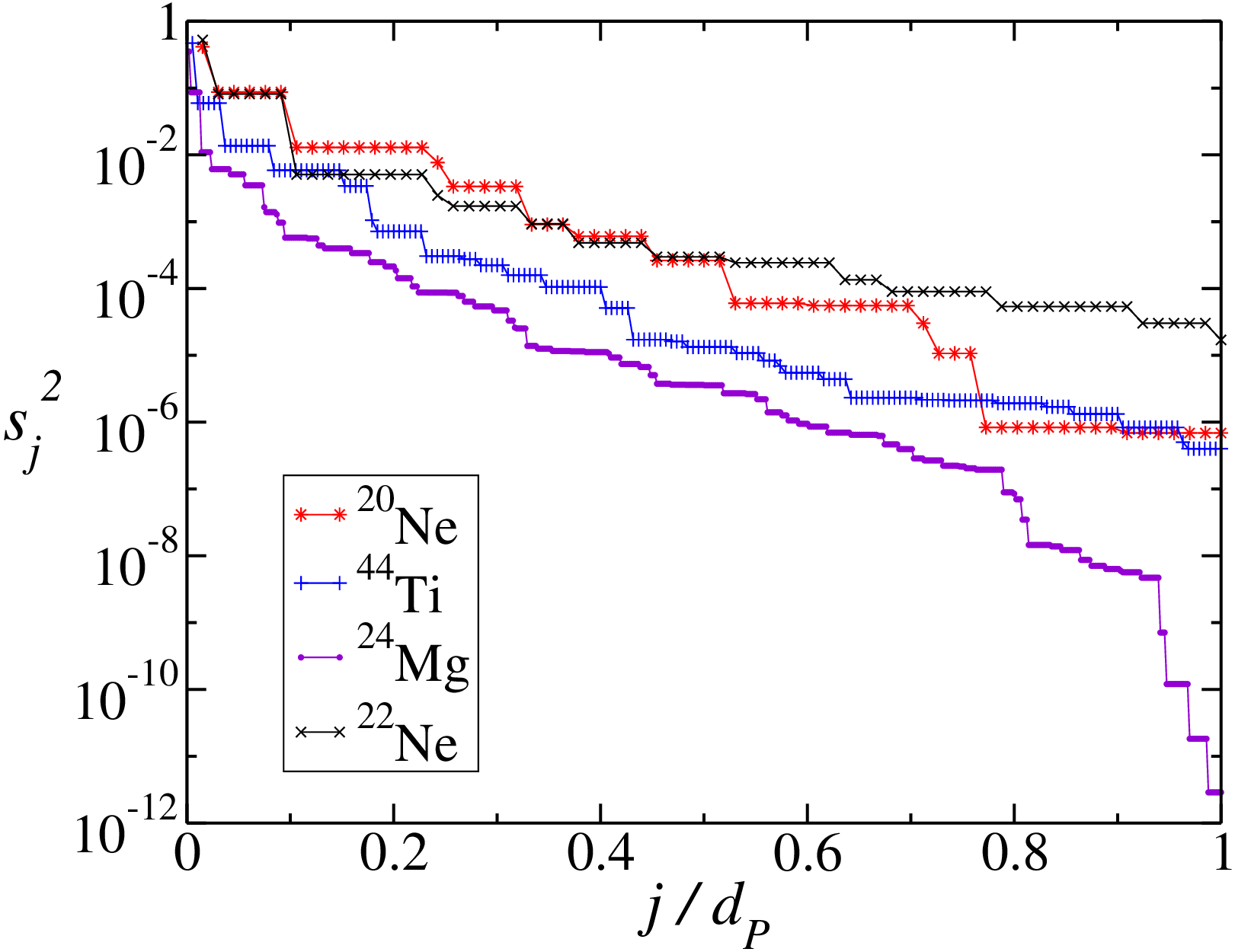}
\caption{\label{fig1}Singular values $s_j^2$ for ground-states of 
${}^{20}$Ne ($\ast$, $d_{\cal P}=66$),
${}^{22}$Ne ($\times$, $d_{\cal P}=66$),
${}^{44}$Ti ($+$, $d_{\cal P}=190$), and
${}^{24}$Mg ($\bullet$, $d_{\cal P}=495$).}
\end{figure}

We divide the set of single-particle orbitals into two subsets, ${\cal
P}$ and ${\cal N}$, respectively. Here ${\cal P}$ represents the set of all
proton orbitals while ${\cal N}$ denotes the set of all neutron orbitals.
We label many-body basis states within each subset as
$\{|\pi_\alpha\rangle,\alpha=1,\ldots,d_{\cal P}\}$ and
$\{|\nu_\alpha\rangle,\alpha=1,\ldots,d_{\cal N}\}$. We expand
a nuclear many-body state $|\Psi\rangle$ in terms
of the proton and neutron states
\be
\label{state}
|\Psi\rangle=\sum_{\alpha}^{d_{\cal P}}\sum_{\beta}^{d_{\cal N}} 
\Psi_{\alpha \beta}
|\pi_\alpha\rangle|\nu_\beta\rangle.
\ee 
This expansion is not unique since the amplitudes $\Psi_{\alpha
\beta}$ depend on the choice of basis states within the two subsets. 
There is, however, a
preferred basis in which the amplitudes $\Psi_{\alpha \beta}$ are
diagonal. This basis is obtained from a singular value
decomposition of the amplitude matrix $\Psi_{\alpha \beta}$, i.e.
$\Psi = U S V^T$, with diagonal $S$ and orthogonal matrices $U$ and $V$.
\be
\label{svd}
|\Psi\rangle=\sum_{j=1}^{\min{\left(d_{\cal P},d_{\cal N}\right)}}
s_j |\tilde{p}_j\rangle|\tilde{n}_j\rangle.
\ee
Here $s_j$ denote the singular values while the correlated 
${\cal P}$-states
$|\tilde{p}_j\rangle$ and the ${\cal N}$-states $|\tilde{n}_j\rangle$
are orthonormal sets of left and right singular vectors, respectively.
The non-negative singular values $s_1\ge s_2\ge s_3 \ldots$ fulfill
$\sum_j s_j^2=1$ due to wave function normalization. It is interesting
to take ground-states of realistic nuclear many-body Hamiltonians and
compute their singular value decomposition in terms of proton and
neutron states. To this purpose we take the amplitude matrix
$\Psi_{\alpha \beta}$ of the ground state (\ref{state}) 
and perform a numerical
singular value decomposition of this matrix. Figure~\ref{fig1} shows
the squares of the singular values for the ground-states of $sd$-shell
nuclei ${}^{20}$Ne, ${}^{22}$Ne and ${}^{24}$Mg (from the USD
interaction \cite{BW}) and for the $pf$-shell nucleus ${}^{44}$Ti
(from the KB3 interaction \cite{KB}). Note that the singular values
decrease rapidly (degeneracies are due to spin/isospin symmetry).
This suggests that a truncation of the factorization (\ref{svd})
should yield accurate approximations to the ground-state. The density
matrix renormalization group (DMRG) \cite{White92}, exploits this
rapid fall-off of singular values in a wave-function
factorization. For first applications of this method to nuclear
structure problems we refer the reader to
Refs.~\cite{Duk01,Duk02,Dimitrova02}.

In this paper, we present an alternative technique that efficiently factorizes
shell-model ground-states and determines the optimal factors for a given
truncation. We make the ansatz
\be
\label{ansatz}
|\psi\rangle=\sum_{j=1}^\Omega |p_j\rangle|n_j\rangle.
\ee
Here, the unknown factors are the ${\cal P}$-states $|p_j\rangle$ and
the ${\cal N}$-states $|n_j\rangle$. The truncation is controlled by
the parameter $\Omega$ which counts the number of desired
factors. Figure~\ref{fig1} suggests that $\Omega\ll\min{\left(d_{\cal
P},d_{\cal N}\right)}$ yields accurate approximations to shell-model
ground-states. This is also the result of our numerical computations
below.

Let $\hat H$ be the nuclear many-body Hamiltonian.
The unknown ${\cal P}$-states $|p_j\rangle$ and ${\cal N}$-states
$|n_j\rangle$ in Eq.~(\ref{ansatz}) are obtained from a variation of the 
energy $E=\langle\psi|\hat{H}|\psi\rangle/\langle\psi|\psi\rangle$, which 
yields ($j=1,\ldots,\Omega$)
\ba
\label{solution}
\sum_{i=1}^\Omega\left(\langle n_j|\hat{H}|n_i\rangle 
  - E\langle n_j|n_i\rangle\right)|p_i\rangle &=& 0,\nonumber\\
\sum_{i=1}^\Omega\left(\langle p_j|\hat{H}|p_i\rangle 
  - E\langle p_j|p_i\rangle\right)|n_i\rangle &=& 0.
\ea
These nonlinear equations are not easy to solve simultaneously. Note
however that for fixed neutron (proton) states the first (second)
set of these equations constitutes a generalized eigenvalue problem for 
the proton (neutron) states. Consider the first set of the 
Eqs.~(\ref{solution}). The operator
$\langle n_i|\hat{H}|n_j\rangle$ acts on ${\cal
P}$-space and is determined by the nuclear structure Hamiltonian
\be
\label{ham}
\hat{H}=\hat{H}_{\cal N} + \hat{H}_{\cal P} +\hat{V}_{\cal PN}, 
\ee
with
\ba
\hat{H}_{\cal N}&=&\sum_n \varepsilon_n \hat{a}_n^\dagger\hat{a}_n
+ {1\over 4}\sum_{n,n',m,m'}v_{n n' m' m}\hat{a}_n^\dagger
\hat{a}^\dagger_{n'}\hat{a}_m\hat{a}_{m'},\nonumber\\
\hat{H}_{\cal P}&=&\sum_p \varepsilon_p \hat{a}_p^\dagger\hat{a}_p
+ {1\over 4}\sum_{p,p',q,q'}v_{p p' q' q}\hat{a}_p^\dagger
\hat{a}^\dagger_{p'}\hat{a}_q\hat{a}_{q'},\\
\hat{V}_{\cal PN}&=& \sum_{p,n,n',p'} v_{p n p' n'}
\hat{a}_p^\dagger\hat{a}^\dagger_{n}\hat{a}_{n'}\hat{a}_{p'}.\nonumber
\ea
Here, we use indices $p,q$ and $m,n$ to refer to proton and neutron
orbitals, respectively. The antisymmetric two-body matrix elements are
denoted as $v_{i j k l}$.

Thus, the ${\cal P}$-space Hamilton operator 
$\langle n_j|\hat{H}|n_i\rangle$ is 
\ba
\label{Hpp}
\langle n_i|\hat{H}|n_j\rangle &=& 
\sum_{p,p'} \left(\sum_{n,n'}v_{p n p' n'}
\langle n_i|\hat{a}^\dagger_{n}\hat{a}_{n'}|n_j\rangle\right)
\hat{a}_p^\dagger\hat{a}_{p'}\nonumber\\
&+& \langle n_i|\hat{H}_{\cal N}|n_j\rangle+
\langle n_i|n_j\rangle\hat{H}_{\cal P}. 
\ea
Note that the neutron-proton interaction $\hat{V}_{\cal PN}$ results into
a one-body proton operator while the neutron Hamiltonian
$\hat{H}_{\cal N}$ yields a constant. This concludes the detailed 
explanation of the first set of equations in Eq.~(\ref{solution}). 
The second set has an identical structure, only the role of neutrons and
protons is reversed.

We solve the coupled set of nonlinear equations (\ref{solution})
iteratively as follows. We choose a random set of initial ${\cal
N}$-states $\{|n_j\rangle, j=1,\ldots,\Omega\}$ and solve the first set
of Eq.~(\ref{solution}) for those ${\cal P}$-states
$\{|p_j\rangle,j=1,\ldots,\Omega\}$ that yield the lowest energy
$E$. We then use this solution for the second set of
Eq.~(\ref{solution}) which yields improved ${\cal N}$-states
$\{|n_j\rangle,j=1,\ldots,\Omega\}$. We iterate this procedure until
the energy $E$ is converged, typically 5-20 times. The advantage of
the factorization method is that the dimensionality of the eigenvalue
problem is $\Omega\times\max{(d_{\cal P},d_{\cal N})}$ while an exact
diagonalization scales like $d_{\cal P}\times d_{\cal N}$.  Before we
present numerical results we note two further developments of the
method. First, we will exploit the non-uniqueness of the ansatz
(\ref{ansatz}) to reduce the generalized eigenvalue problem to a
standard eigenvalue problem. Second, we will use the rotational
symmetry of the shell-model Hamiltonian (\ref{ham}) to give a
$m$-scheme formulation of the factorization method.

To reduce the Eqs.~({\ref{solution}) to a standard eigenvalue problem
we choose an orthonormal set of initial ${\cal N}$-states
$\{|n_j\rangle,j=1,\ldots,\Omega\}$ as input to the first eigenvalue
problem. This reduces the ``overlap'' matrix $\langle n_j|n_i\rangle$
to a unit matrix, and we solve a standard eigenvalue problem to
obtain the ${\cal P}$-states $\{|p_j\rangle,j=1,\ldots,\Omega\}$. The
resulting ${\cal P}$-states will not be orthogonal in general. Their
coefficient matrix $C$ with elements $C_{\alpha j}\equiv \langle
\pi_\alpha | p_j\rangle$ may, however, be factorized in a singular
value decomposition as $C=U D V^T$. Here $D$ denotes a diagonal
$\Omega\times\Omega$ matrix while $U$ is a (column) orthogonal
$d_{\cal P} \times\Omega$ matrix, and $V$ is a orthogonal
$\Omega\times\Omega$ matrix. The transformed states
($j=1,\ldots,\Omega$)
\bas
|p_j'\rangle&=&\sum_{\alpha=1}^{d_{\cal P}} U_{\alpha,j}|\pi_\alpha\rangle, 
\nonumber\\ 
|n_j'\rangle&=&\sum_{i=1}^\Omega V_{ij} |n_i\rangle
\eas
are orthonormal in ${\cal P}$-space and ${\cal N}$-space, respectively, and
fulfill 
\be
|\psi\rangle=\sum_{j=1}^\Omega |p_j\rangle|n_j\rangle = 
\sum_{j=1}^\Omega D_{jj}|p_j'\rangle|n_j'\rangle.
\ee
We then input the new ${\cal P}$-states
$\{|p_j'\rangle,j=1,\ldots,\Omega\}$ to the second set in
Eq.~(\ref{solution}), which poses a standard eigenvalue problem for
the ${\cal N}$-states. The transformed neutron states
$D_{jj}|n'_j\rangle$ are good starting vectors for the Lanczos
iteration. We iterate the whole procedure until the ground-state
energy converges to an appropriate level of accuracy. 
Note that the singular value decomposition is very
inexpensive compared to the diagonalization. This procedure also has
the advantage that it yields the singular values $D_{jj}$ of the
ground-state factorization.

\begin{table*}[t]
\begin{ruledtabular}
\begin{tabular}{|l|rr|rr|rr|rr|}
        &\multicolumn{2}{c|}{${}^{44}$Ti}&\multicolumn{2}{c|}{${}^{48}$Cr}&\multicolumn{2}{c|}{${}^{52}$Fe}&\multicolumn{2}{c|}{${}^{56}$Ni}\\\hline  
    &$E_0$ [MeV]& $d$   & $E_0$ [MeV]&  $d$   &$E_0$ [MeV]&$d$&$E_0$ [MeV]&$d$\\\hline
Factorization  &   -10.74   & 190     & -29.55   &  4,845     &  -50.41 & 38,760& -76.23&125,970\\
$m$-scheme Fact.& -12.74   & 190     & -31.06   &  4,845     &  -52.03 & 38,760 & -77.1 &125,970\\
Exact   &   -13.88   & 4,000   & -32.95   &  1,963,461 & -54.27 & 109,954,620 & -78.46 & 1, 087,455,228\\
\end{tabular}
\end{ruledtabular}
\caption\protect{\label{tab1}
Results for ground-state energies $E_0$ of $pf$-shell nuclei obtained
from the factorization with just one state and the most severe
$m$-scheme factorization compared to the exact results from
Ref.~\cite{Nathan}. The $m$-scheme dimension of the corresponding
eigenvalue problem, $d$, is also listed.}
\end{table*}

To include the axial symmetry (``$m$-scheme'') into the factorization
we consider only products of proton and neutron states that have angular
momentum projection $J_z=0$. To this 
purpose we modify the ansatz (\ref{ansatz}) as
\be
\label{mscheme}
|\psi\rangle=\sum_m\sum_{j=1}^{\Omega_m} |p_j^{(m)}\rangle
|n_j^{(-m)}\rangle,
\ee
where $|p_j^{(m)}\rangle, j=1,\ldots,\Omega_m$ ($|n_j^{(-m)}\rangle,
j=1,\ldots,\Omega_{m}$) denote many-proton states (many-neutron
states) with angular momentum projection $J_z=m$ ($J_z=-m$), and the sum
over $m$ runs over all value of $J_z$ that are realized.  The ansatz
(\ref{mscheme}) again leads to a generalized eigenvalue problem
similar to Eq.~(\ref{solution}), which we also cast into a
standard eigenvalue problem by maintaining orthogonality between ${\cal
P}$-states and between the ${\cal N}$-states through singular value
decompositions. The number of factors used in the $m$-scheme factorization
is given by the parameters $\Omega_m$. In the following, we use
\be
\label{frac}
\Omega_m(\alpha)=\max{\left(1, \alpha\Omega_m^{\rm max}\right)},
\ee
where $\Omega_m^{\rm max}$ is the maximal dimension of the
corresponding subspace. For $\alpha=0$ the most severe truncation is
obtained while $\alpha=1$ leads to an eigenvalue problem with the same
dimension as an exact diagonalization in $m$-scheme. In the latter
case, the first iteration (i.e. solving the first set of
Eq.~(\ref{solution}) for the proton states and using random,
orthonormal neutron states as input) leads to the exact solution. This
is due to the non-uniqueness of the ansatz (\ref{mscheme}). Note that
other abelian symmetries and point symmetries can implemented in a
similar manner as the axial symmetry. This flexibility is important
since it allows us to consider factorizations that differ from the
neutron-proton factorization proposed here. In neutron rich nuclei,
for instance, there might be a large imbalance between the sizes of
the proton and neutron space. In such a situation one 
might consider a ``mixed'' factorization where
part of the neutron orbitals and all proton orbitals constitute one 
factor space while the other factor space consists of the remaining 
neutron orbitals. Note also that factorization method proposed here
is not restricted to nuclear structure problems but could well be applied
to problems in quantum chemistry or condensed matter.

We turn to numerical tests of the factorization method. We solve the
eigenvalue equations with the sparse matrix solver {\sc arpack}
\cite{arpack}. In a first step we set $\Omega=1$ in the ansatz
(\ref{ansatz}) and compare the most severely truncated factorization
with exact diagonalization. The factorization requires us to
iteratively solve eigenvalue problems with rather low dimension
$d_{\cal P}$ and $d_{\cal N}$, respectively. Table~\ref{tab1} shows
the results for the $pf$-shell nuclei ${}^{44}$Ti, ${}^{48}$Cr,
${}^{52}$Fe, and ${}^{56}$Ni. The results from the severely truncated
factorization deviate about 2-4 MeV from the exact results (from
Ref.~\cite{Nathan}). For comparison, we also list the dimension of the
eigenvalue problem. It is also interesting to use the $m$-scheme
factorization and set $\Omega_m=1$ in Eq.~(\ref{mscheme}). This
corresponds to $\alpha=0$ in Eq.~(\ref{frac}) and yields a problem of
equal dimension as the factorization (\ref{ansatz}) with truncation
$\Omega=1$. Results are also listed in Table~\ref{tab1}. The advantage
of the $m$-scheme factorization is clearly visible as the
corresponding results typically differ only 1-2 MeV from the exact
results. This makes this severe truncation particularly interesting to
compute good starting vectors for large-scale nuclear structure
problems.  After these encouraging results we will in the following
include more factors to obtain more accurate ground-state
factorizations.

We use the $m$-scheme factorization to compute accurate approximations
to ground-state energies. As a first test case we consider the
$sd$-shell nucleus ${}^{24}$Mg with the USD interaction \cite{BW}.
The exact diagonalization yields the ground-states energy $E_0=-87.08$
MeV and has $m$-scheme dimension $d_{\rm max}=28503$.  We solve the
eigenvalue problem (\ref{solution}) for the ground state but also
record a few excited energy solutions.  The hollow data points in
Fig.~\ref{fig2} show the resulting excitation spectrum versus the
dimension of the eigenvalue problem relative to the dimension of the
full $m$-scheme diagonalization, $d/d_{\rm max}$.  The ground-state
converges very fast while the excited states converge somewhat slower
toward the exact results.  To obtain fast converging results for the
excited states, we directly solve the eigenvalue problem
(\ref{solution}) for excited energies.  The full data points in
Fig.~\ref{fig2} show the results of such calculations for the first
and second excited state, respectively.
  
\begin{figure}[t]
\vskip 0.3cm
\includegraphics[width=0.45\textwidth]{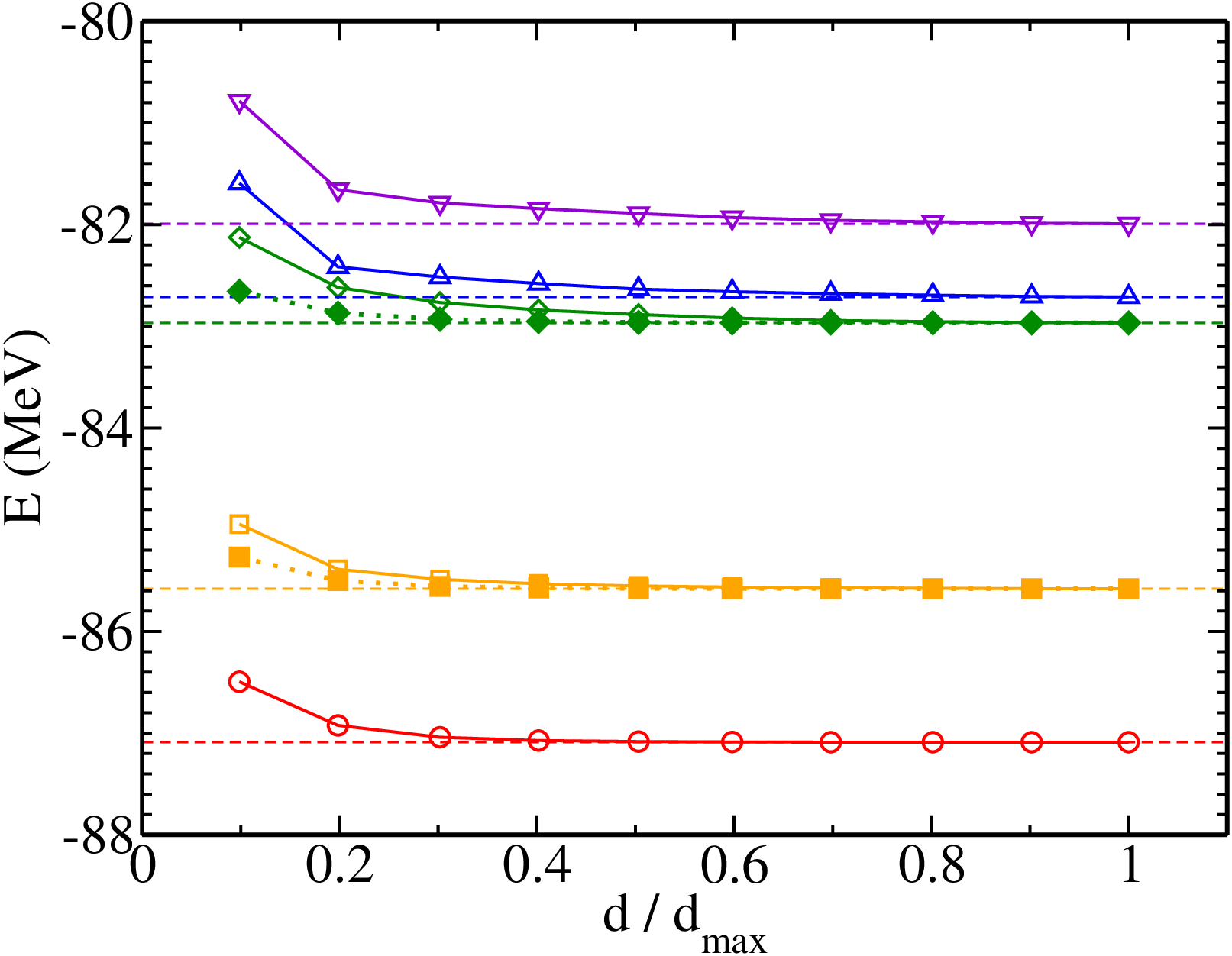}
\caption{\label{fig2}Low-energy spectrum for ${}^{24}$Mg (USD interaction)
versus the relative dimension
of the eigenvalue problem. The dashed lines are the exact results, the
open data points result from targeting the ground-state, and
the filled data points result from directly targeting the 1$^{st}$ and
2$^{nd}$ excited state, respectively.}
\end{figure}

Figure~\ref{fig3} shows the squared overlaps of the factorization
results with the exact solution versus the dimension $d$ of the
eigenvalue problem. Excellent results are obtained for the ground
state (which was directly targeted) and for the excited states. The
inset of Fig.~\ref{fig3} shows that the angular momenta of the
low-lying states are accurately reproduced even at severe
truncations. This indicates that transition matrix elements can
also be calculated very accurately. The results of Fig.~\ref{fig2} and
Fig.~\ref{fig3} clearly demonstrate the efficiency and accuracy of the
factorization.

\begin{figure}[b]
\vskip 0.3cm
\includegraphics[width=0.45\textwidth]{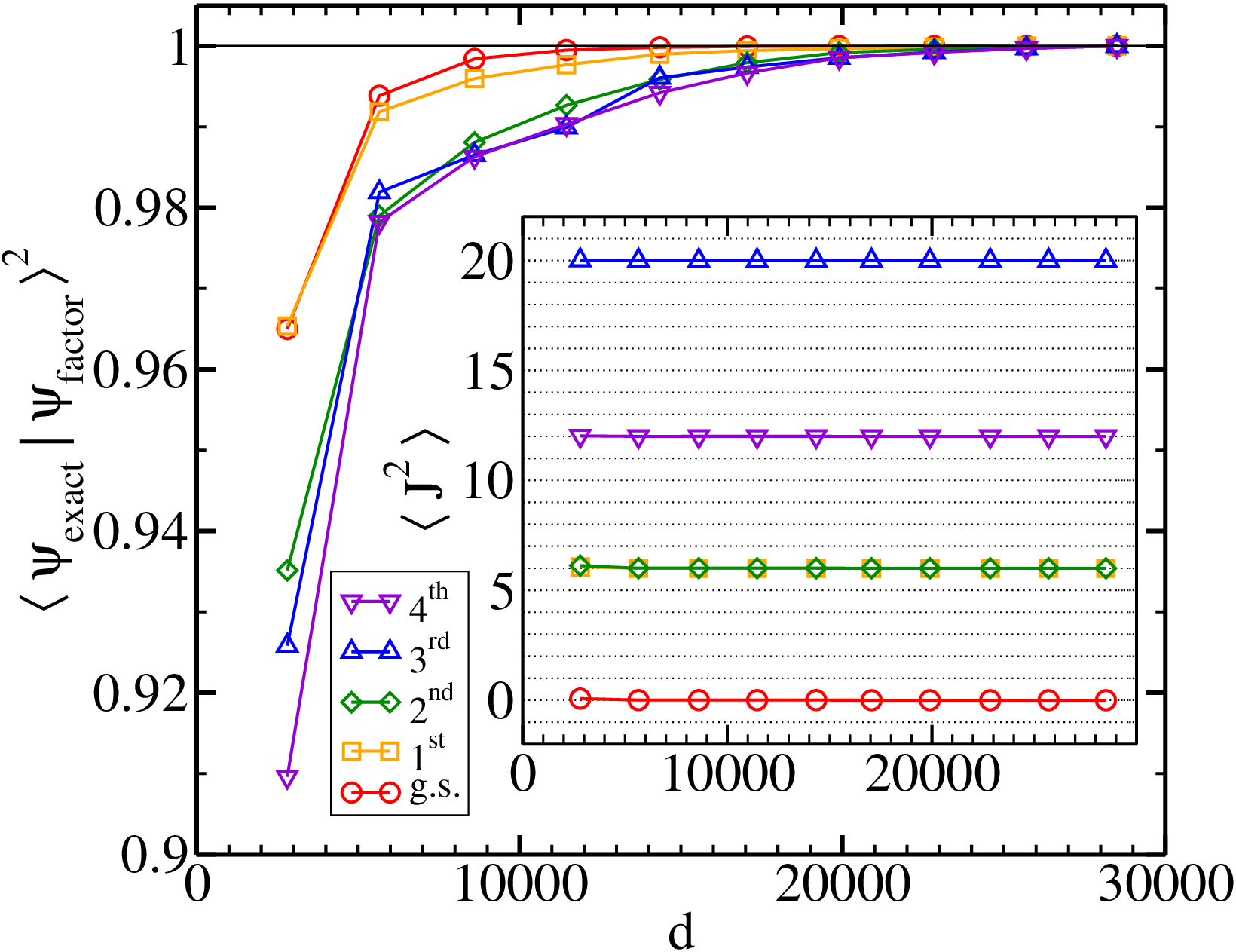}
\caption{\label{fig3} Squared wave function overlaps with the exact
results for the low-lying states in ${}^{24}$Mg (USD interaction)
versus the dimension of the eigenvalue problem. Inset: Corresponding
angular momentum expectation values. The results are obtained from
targeting the ground state.}
\end{figure}

For $pf$-shell nuclei we use the KB3 interaction \cite{KB} and compute
the ground-states of $^{56}$Ni and ${}^{48}$Cr. The respective
$m$-scheme dimensions are $d_{\rm max}=1.087\times 10^9$ and $d_{\rm
max}=1.96\times 10^6$. Figure~\ref{fig4} shows the ground-state
energies for $^{56}$Ni and ${}^{48}$Cr (inset) plotted versus the
relative dimension $d/d_{\rm max}$ of the respective eigenvalue
problem. We also show an exponential fit of the form $E(d/d_{\rm
max})=E_0+b\exp{(-cd/d_{\rm max})}$ to the rightmost seven and six
data points, respectively.  For $^{56}$Ni (${}^{48}$Cr) the resulting
ground-state energy is only 100 keV (30keV) above the exact result
while the dimension of the eigenvalue problem is dramatically reduced
by a factor 1000 (factor 12). Figure ~\ref{fig5} shows the excitation
spectrum relative to the ground-state. Our method reproduces the level
spacings of the two lowest excitations very accurately even at the
most severe truncation, while the spacings to the higher levels are
about 300 keV too large. Quantum numbers are reproduced accurately for
$d/d_{\rm max}\agt 0.05$. Considering the modest size of the
eigenvalue problem we solved, these are very good results.  We also
compared the $m$-scheme factorization with a particle-hole
calculation. The results are listed in Table~\ref{tab2} and clearly
demonstrate the fast convergence of the factorization method.

\begin{figure}[t]
\vskip 0.3cm
\includegraphics[width=0.45\textwidth]{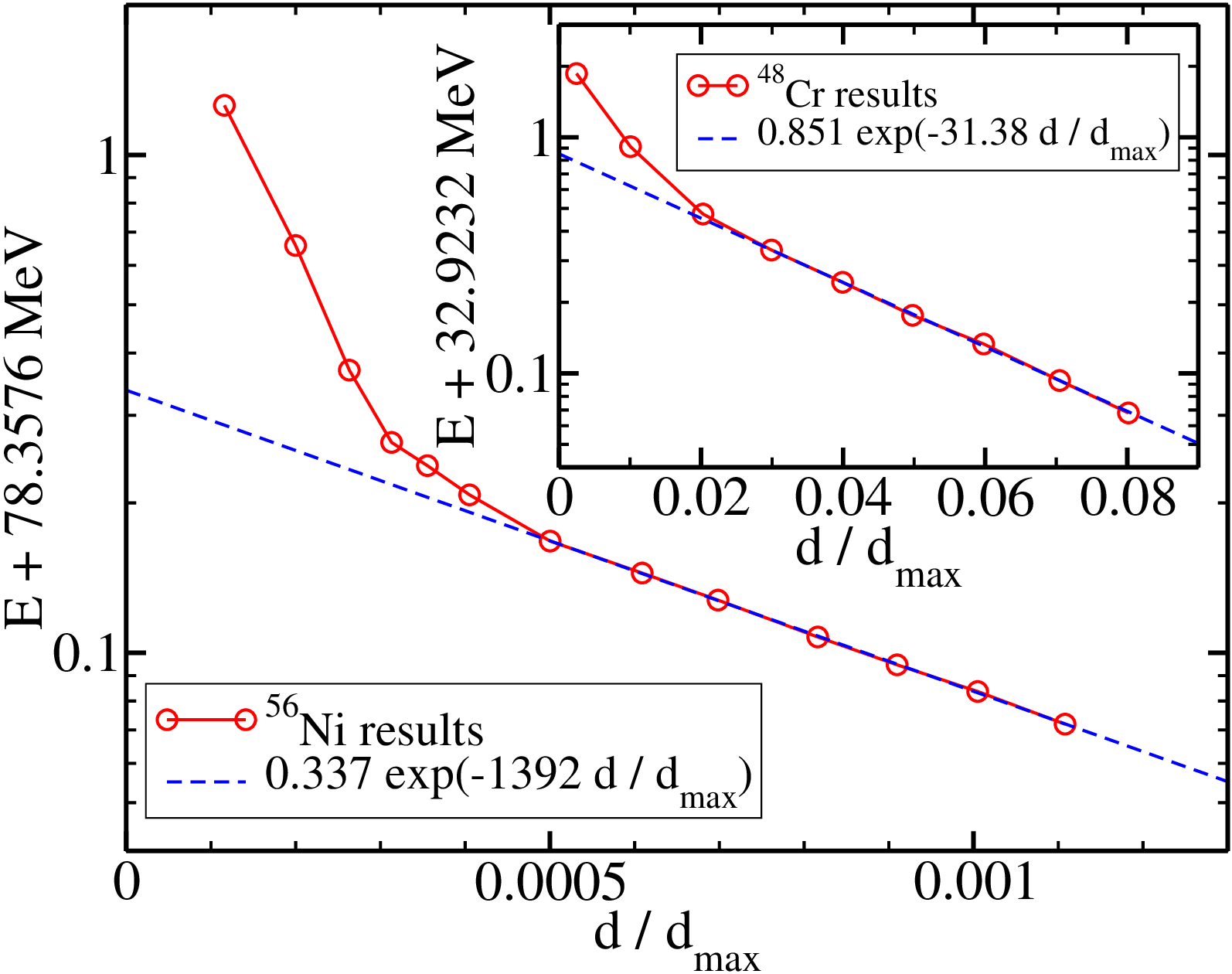}
\caption{\label{fig4}Ground-state energy $E$ versus the dimension $d$
of the eigenvalue problem relative to the $m$-scheme dimension 
$d_{\rm max}=1.087\times 10^9$ for $^{56}$Ni. Inset: Similar plot for
${}^{48}$Cr ($d_{\rm max}=1.96\times 10^6$).} 
\end{figure}

Let us also compare the ground-state factorization presented in this
work to three related methods. The DMRG also bases its truncation on
the singular values \cite{White92}. Recent applications to simple
nuclear structure problems were very successful \cite{Duk01,Duk02},
but the application to realistic nuclear structure problems seems more
difficult as the convergence is very slow \cite{Dimitrova02}.
Recently, Andreozzi and Porrini \cite{AP} approximated the shell-model
ground-state by products of eigenstates of the proton-proton and
neutron-neutron Hamiltonian. This method yields quite good results for
energy spacings of low-lying shell-model states. However, the
generation of the eigenstates might become impractical for the large
proton and neutron spaces. A third related method is the Exponential
Convergence Method (ECM) \cite{Horoi94,Horoi99,Horoi02,Horoi03}. A
direct comparison is not easy since the FPD6 interaction is used for
$pf$-shell nuclei, and since ECM results are plotted versus
$JT$-coupled dimension of the truncated space.  However, it seems that
the convergence of our method is considerably faster. For $^{48}$Cr,
for instance, our rate of exponential convergence is $c\approx -31.38$
(See Fig.~\ref{fig4}), which is about a factor eight larger than what
is reported for the ECM in Fig.~1 of Ref.\cite{Horoi02}. For
$^{56}$Ni, our exponential rate is about a factor 200 larger than the
ECM rate \cite{HoroiPC}, and our identification of the exponential
region requires a $m$-scheme dimension $d\approx 10^6$ (See
Fig.~\ref{fig4}) while the ECM requires an $m$-scheme dimension of 4-5
million \cite{HoroiPC}.

\begin{figure}[t]
\vskip 0.3cm
\includegraphics[width=0.45\textwidth]{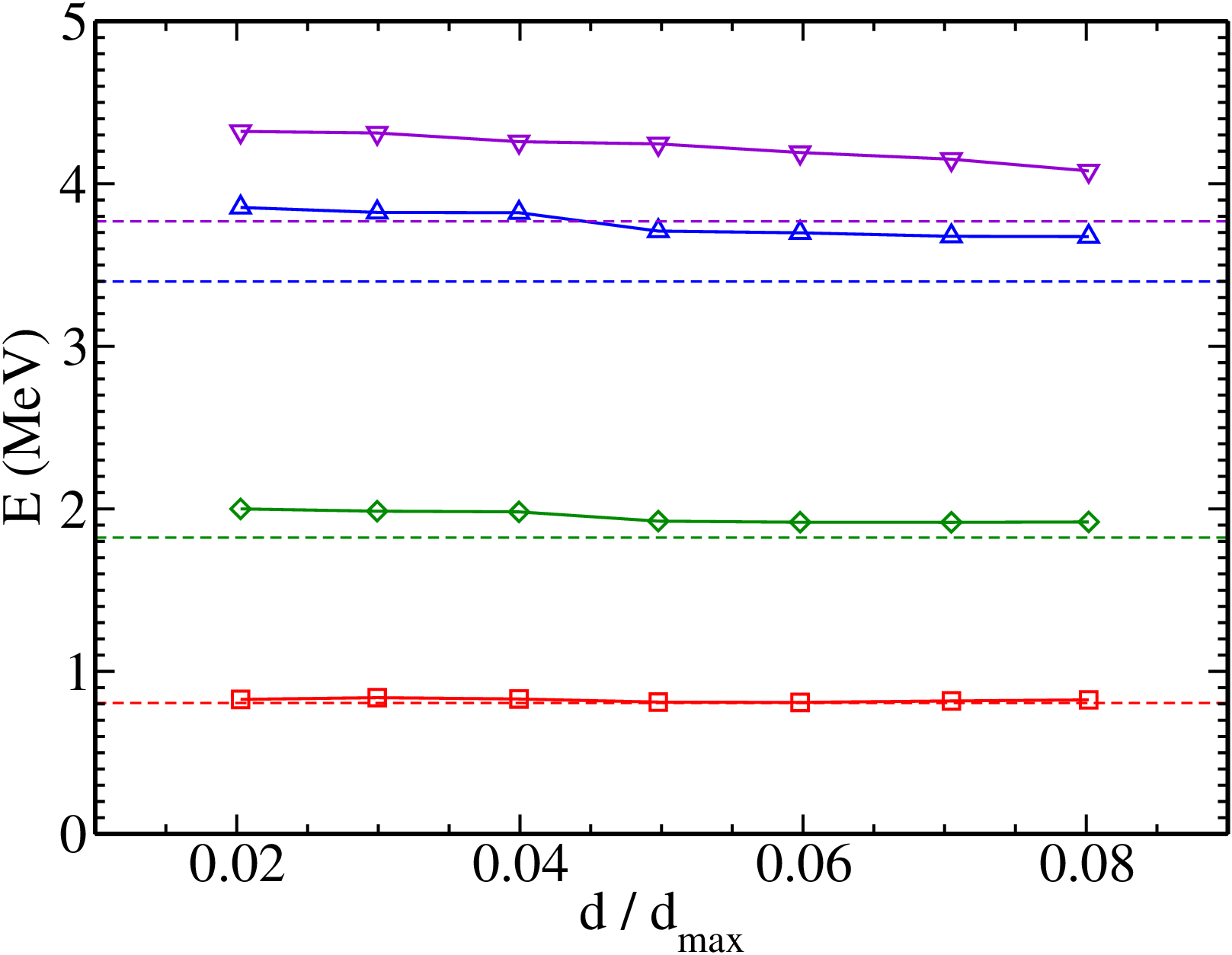}
\caption{\label{fig5}Level spacings of the lowest excitations 
with respect to the ground-state
of the $pf$-shell nucleus ${}^{48}$Cr (KB3 interaction) plotted versus
the dimension $d$ of the eigenvalue problem relative to the $m$-scheme
dimension $d_{\rm max}$. The dashed lines are the exact results.}
\end{figure}

In summary, we proposed a new method that factorizes ground-states of
realistic nuclear structure Hamiltonians. The optimal factors are
derived from a variational principle and are the solution of rather
low-dimensional eigenvalue problems. The approximated states and
energies converge exponentially quickly as more factors are included,
and quantum numbers are accurately reproduced.  Computations for
$sd$-shell and $pf$-shell nuclei show that highly accurate
approximations may result from eigenvalue problems whose dimensions are
reduced by orders of magnitude.

\begin{table}[t]
\begin{ruledtabular}
\begin{tabular}{|c|r||c|c|r|}
\multicolumn{2}{|c||}{$m$-scheme fact.}&\multicolumn{3}{c|}{p-h approach}\\\hline
  $E_0$ [MeV]  &  $d$     &         & $E_0$ [MeV]    & $d$       \\\hline
   -31.06      &   4,845  & 2p-2h   & -31.11         &    62,220 \\
   -32.68      &  78,407  & 4p-4h   & -32.62         &   736,546 \\
   -32.83      & 138,386  & 5p-5h   & -32.83         & 1,328,992 \\
\end{tabular}
\end{ruledtabular}
\caption\protect{\label{tab2}
Comparison of $m$-scheme factorization method with a particle-hole
calculation for ${}^{48}$Cr (KB3 interaction). $d$ denotes the dimension
of the corresponding eigenvalue problem.}
\end{table}

The authors thank G. Stoitcheva for useful discussions and help in the
uncoupling of matrix elements, and acknowledge communications with
M. Horoi.  This research used resources of the Center for
Computational Sciences at Oak Ridge National Laboratory, and was
supported in part by the U.S. Department of Energy under Contract
Nos.\ DE-FG02-96ER40963 (University of Tennessee) and
DE-AC05-00OR22725 with UT-Battelle, LLC (Oak Ridge National
Laboratory).

\end{document}